\documentclass[twocolumn,secnumarabic,amssymb,superscriptaddress, nobibnotes, aps, prl,reprint]{revtex4-2}

\usepackage{pdfpages}
\usepackage{xcolor}
\usepackage{float}
\usepackage{lipsum}
\usepackage{graphicx}
\usepackage{dcolumn}
\usepackage{bm}
\usepackage[normalem]{ulem}
\usepackage{soul}
\usepackage{dblfloatfix}
\usepackage{amsmath}
\usepackage{pgffor}

\makeatletter
\AtBeginDocument{\let\LS@rot\@undefined}
\makeatother

\begin{document}

\title{Optical flux pump in the quantum Hall regime}

\author{Bin Cao}
\affiliation{Joint Quantum Institute, NIST/University of Maryland, College Park, Maryland 20742, USA}
\author{Tobias Grass}
\affiliation{ICFO-Institut de Ciencies Fotoniques, The Barcelona Institute of Science and Technology, Castelldefels (Barcelona) 08860, Spain}
\author{Glenn Solomon}
\affiliation{Joint Quantum Institute, NIST/University of Maryland, College Park, Maryland 20742, USA}
\author{Mohammad Hafezi}
\affiliation{Joint Quantum Institute, NIST/University of Maryland, College Park, Maryland 20742, USA}
\affiliation{IREAP, University of Maryland,
College Park, Maryland 20742, USA}


\newcommand\rd[1]{{\color{red}#1}}
\newcommand\cross[1]{{\color{red}\sout{#1}}}

\begin{abstract}
A seminal \emph{gedankenexperiment} by Laughlin describes the charge transport in quantum Hall systems \emph{via} the pumping of flux. Here, we propose an optical scheme which probes and manipulates quantum Hall systems in a similar way: When light containing orbital angular momentum interacts with electronic Landau levels, it acts as a flux pump which radially moves the electrons through the sample. We investigate this effect for a graphene system with Corbino geometry, and calculate the radial current in the absence of any electric potential bias. Remarkably, the current is robust against the disorder which is consistent with the lattice symmetry, and in the weak excitation limit, the current shows a power-law scaling with intensity characterized by the novel exponent $2/3$.
\end{abstract}

\maketitle
\textit{Introduction.} Multipole transitions beyond the dipole approximation apply when the Bohr radius of the quantum state is larger or comparable to the excitation wavelength. This is rarely the case for atoms or quantum dots \cite{andersen2011strongly,schmiegelow2016transfer,Afanasev2018,de2020photoelectric,allen1992orbital,franke2017optical}. However, in the quantum Hall regime, wavefunctions can be extended to a length scale comparable to optical wavelengths~\cite{feldman2016observation,ghahari2017off,deprez2020tunable,ronen2020aharonov,wei2017mach,mills2019dirac}, and the coherence is topologically protected against dephasing~\cite{ronen2020aharonov,zhang2005experimental}. Consequently, multipole transitions become possible~\cite{gullans2017high,grass2018optical,takahashi2018landau,takahashi2019selection}. Specifically, if the optical field has an orbital angular momentum (OAM)~\cite{allen1992orbital,franke2008advances}, these transitions can transfer angular momentum from photons to electrons, and an interesting interplay between topological properties of electrons and photons may be observed~\cite{bliokh2012electron,mciver2020light,kim2020optical}. Similar effects also exist in synthetic quantum Hall system made of Rydberg polaritons~\cite{ivanov2018adiabatic}.
Outside the quantum Hall regime, incoherent multipole interactions between light and condensed matter systems have been experimentally studied~\cite{ji2020photocurrent,sederberg2020vectorized,simbulan2020twisted,sanvitto2010persistent,kwon2019direct}; and many theoretical efforts have been made~\cite{quinteiro2009electric,farias2013photoexcitation,watzel2012photovoltaic}. However, there has been no observation of coherent multipole interaction with quantum Hall states, to the best of our knowledge. In this context, disorder may play an important role, as it mixes eigenstates of angular momentum, but previous studies have largely ignored its effect~\cite{grass2018optical,takahashi2018landau,takahashi2019selection}.
 
 \begin{figure}
\centering
\includegraphics[scale=1]{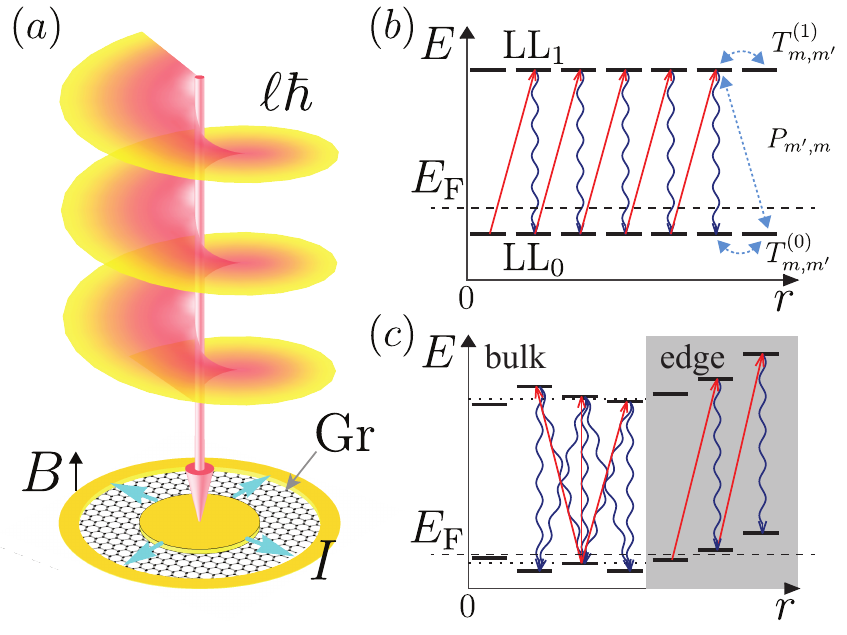}
\caption{(a) The illustration of the proposed setup. The Corbino-structured sample is concentric with the OAM beam. The OAM-induced current is measured between the inner and outer electric contacts. (b) In a pristine system, the orbitals in each LL are degenerate in energy. The optical selection (red arrows) leads to an increase of OAM (for positive $\ell$) and phonon relaxations (blue arrows) maintain the OAM. This equivalently leads to a directional transport of electrons. (c) In a system with disorder and confinement, the LLs are broadened in the bulk while the energy of the edge orbitals (shaded region) rises quickly. Only the edge states couple and relax pair-wisely while the disordered bulk eigenstate may couple and relax to more than one eigenstate.}
\label{Fig1}
\end{figure}

Here, to observe such a topological interplay, we theoretically study the interaction between light with OAM and a graphene device with Corbino geometry in the quantum Hall regime, see Fig.\ref{Fig1}(a). We focus on the radial dynamics of the electrons in Landau levels (LLs) upon illumination of light with non-zero OAM and propose an OAM-induced photocurrent measurement. We solve the Bloch equations incorporating the optical coupling, acoustic phonon relaxation, potential disorders, effective boundaries and Pauli blockade. In particular, we first consider an ideal system without disorder which simplifies to a translationally invariant model which can be analytically solved. This yields an expression for the OAM-induced current, which scales with pump intensity to the power of 2/3. Second, we take into account short- and long-range disorders and boundary effects without intervalley mixings, and numerically solve the Bloch equations. The results show that a radial current is generated as a result of electrons moving outward or inward between orbitals upon absorbing a photon with OAM. The current's direction and amplitude is determined by the OAM of the light with respect to the magnetic field. Remarkably, this is similar to the Laughlin pump~\cite{laughlin1981quantized} where a magnetic flux induces a spectral flow of the electrons' OAM, whereas in our scheme the flux is replaced by a light beam with a phase winding. We find that the current is reduced for larger disorder strength. However, we can recover the current through applying a voltage bias. Finally, we show that the numerical simulated current matches the scalings predicted analytically.

\textit{Landau levels in graphene.} The system considered is a Corbino-shape graphene device~\cite{zeng2019high,li2019pairing} under a strong out-of-plane magnetic field $B$, as shown in Fig.~\ref{Fig1}(a). The magnetic field quenches the kinetic energies of the electrons and their states manifest as LLs. Here we consider two LLs, the zeroth LL ($\mathrm{LL_0}$) and the first LL ($\mathrm{LL_1}$). Under an achievable high magnetic field of 15~T, for example, the frequency corresponding to the transition between the two LLs is around 35 THz which lies in the mid-infared optical regime. In graphene, very few other transitions match the same energy due to the inharmonic level spacing, and only $\mathrm{LL_0}$ to $\mathrm{LL_1}$ is allowed when the Fermi level is set in between them. We refer to transitions between different LLs as interband, while those among orbitals inside a LL as intraband. We ignore carrier-carrier interaction since the time scales make the Coulomb interaction irrelevant as we will show later. Also, we limit our discussion to the $K$ valley and ignore the spin degree of freedom. Without any disorder, the spinor wavefunctions for $\mathrm{LL}_0$ and $\mathrm{LL}_1$ in the $K$ valley of graphene are given as \cite{goerbig2009quantum},
 \begin{equation}
     \bar{\Psi}_{0,\bar{m}}=\begin{pmatrix}
    0  \\
    |0,\bar{m}\rangle
    \end{pmatrix},
    \bar{\Psi}_{1,\bar{m}}=\frac{1}{\sqrt{2}}\begin{pmatrix}
    |0,\bar{m}\rangle  \\
    |1,\bar{m}\rangle
    \end{pmatrix}.
    \label{clean_wf}
 \end{equation}
 There are two quantum numbers for $\bar{\Psi}_{s,\bar{m}}$, the LL index $s\in \{0,1\}$ and the orbital index $\bar{m}\in \mathbb{Z^+}$, which corresponds to OAM in the chosen symmetric gauge. In real space, each orbital $\Psi_{s,\bar{m}}$ looks like a circular ring \cite{feldman2016observation} and the radius of the ring increases as $r_{\bar{m}}=\sqrt{2\bar{m}}l_{\textrm{c}}$, where $l_{\textrm{c}}$ is the magnetic length. Without disorder, energies of these orbitals in each LL are degenerate, as shown in Fig.~\ref{Fig1}(b).
 
 To account for disorder, we include a disorder potential consistent with the symmetries of the lattice $V_{\textrm{dis}}=\gamma(u_0(r,\theta) I+ \mathbf{u}(r,\theta)\cdot \mathbf{\sigma})$ where $\gamma$ is the strength of the disorder; $\gamma u_0(r,\theta)$ represents the long-range disorder arising from \emph{e.g.} charge impurities and $\gamma \mathbf{u}(r,\theta)$ is the short range disorder associated with \emph{e.g.} defects, variations of sub-lattice potentials ($u_z$) and tunneling rates ($u_x$, $u_y$)~\cite{sarma2011electronic,gullans2017high}. Inter-valley scatterings~\cite{beenakker2008colloquium} are not considered in this valley-polarized model and corresponding inter-valley effects may not be compatible. To account for boundary effects, we include a confinement potential $V_{\textrm{cf}}=V_{\textrm{c}}\mathcal{H}(r-r_{\textrm{max}})$, where $\mathcal{H}(\cdot)$ is the Heaviside step function. This modeling of confinement, without intervalley scattering, applies to \emph{e.g.} electrostatically defined edges~\cite{li2018valley,ronen2020aharonov} or zigzag edges~\cite{beenakker2008colloquium}. We diagonalize the potential for $\mathrm{LL_0}$ and $\mathrm{LL_1}$ individually, and obtain the disordered LL eigenstates,
\begin{equation}
    \Psi_{0,m}=\sum_{\bar{m}} c^m_{0,\bar{m}} \bar{\Psi}_{0,\bar{m}},  \quad
    \Psi_{1,m}=\sum_{\bar{m}} c^m_{1,\bar{m}} \bar{\Psi}_{1,\bar{m}}.
    \label{disordered_wf}
 \end{equation}
Here we assume $\gamma\ll \delta$, where $\delta$ is the cyclotron energy (the LL gap), and thus the two LLs do not mix. Because of the disorder, $m$ does not represent OAM anymore, but numerates the orbitals with respect to their energy, as the LL degeneracy has been lifted. We truncate our system size such that $m<m^*$, where $m^*$ is the maximum index of the possibly occupied orbitals in our simulation, determined by the size of the sample through $r_{\mathrm{max}}=\sqrt{2m^*}l_c$ assuming that the coherence length is larger than the system size. With $V_{\textrm{dis}}$ and $V_{\textrm{cf}}$, the orbitals in the bulk will give rise to LL broadening, while the orbitals on the physical edges increase energy with $m$ (Fig.~\ref{Fig1}(c)).

\textit{Light-matter interaction.} We illuminate the sample with a laser beam which is in resonance with the interband transition between $\mathrm{LL}_0$ and $\mathrm{LL}_1$. The beam is concentric with the center of the sample and may contain a non-zero OAM, as shown in Fig.~\ref{Fig1}(a). See \footnote{See Supplemental Material for details. The Supplemental Material includes references to~\cite{picon2010transferring,wendler2014carrier,scully1999quantum,kim2020edge,abanin2007charge,goerbig2009quantum,chklovskii1992electrostatics, gutierrez2018interaction,wendler2015ultrafast,schiffrin2013optical,sederberg2020vectorized}.} for the scenarios where the beam is partially blocked or shifted away from the center.

The light-matter interaction is obtained with the minimal coupling $\bm{p} \rightarrow \bm{p}-e\mathbf{A}$,
\begin{equation}
    H_{\textrm{I}}(t)=ev_{\rm{F}}\mathbf{A}(t)\cdot\sigma,
\end{equation}
where $v_{\rm{F}}$ is the Fermi velocity, $\mathbf{A}(t)$ is the vector potential of light and it can be expressed as,
\begin{eqnarray}
    & \mathbf{A}(t) = \mathbf{A}_0(r,\theta)e^{-i\omega t}+\mathbf{A}^*_0(r,\theta)e^{i\omega t},\\
   & \mathbf{A}_0(r,\theta) = A(r)e^{i\ell\theta}\mathbf{p} .
    \label{A(t)}
\end{eqnarray}
Here, $A(r)$ is the mode of the light, which can be the Bessel mode \cite{durnin1987diffraction}, or the Laguerre-Gauss mode~\cite{allen1992orbital}; $\mathbf{p}$ is the in-plane polarization of the field. The twisted phase term $e^{i\ell\theta}$ represents the OAM carried by the light, and $\ell$ counts the OAM.

In the interaction picture, \emph{i.e.} after a unitary transformation which describes the system in a rotating frame with frequency $\omega$, the time-dependence is removed from the Hamiltonian, and we have the light-matter interaction Hamiltonian in graphene as, $H_{\textrm{I}}=ev_{\textrm{F}}A(r)(e^{i\ell\theta}\sigma_{-}+e^{-i\ell\theta}\sigma_{+})$, where we assume that the field is right-circular polarized, \emph{i.e.} $\mathbf{p}=\mathbf{p}_+$. Only the off-diagonal terms are non-zero and they correspond to interband optical transitions between $\mathrm{LL}_0$ and $\mathrm{LL}_1$~\cite{Note1}.

\textit{Bloch equations.} We define the annihilation and creation operators $a_{m}$, $a^\dagger_{m}$ for electrons in orbital $m$ in $\mathrm{LL}_0$, and $b_{m}$, $b_m^\dagger$ for $\mathrm{LL}_1$. They satisfy, $\{a_{m},a^\dagger_{m'}\}=\{b_{m},b^\dagger_{m'}\}=\delta_{m,m'}$. We can rewrite the light-matter interaction as,
\begin{equation}
    H_{\textrm{I}}=\sum_{m,m'} \Omega_{m',m} b^\dagger_{m'} a_{m}+ \Omega^*_{m',m} a^\dagger_{m} b_{m'}
\end{equation}
The Rabi frequency for each pair of orbitals is obtained as, $ \Omega_{m',m}= \langle \Psi_{0,m}|H_{\textrm{I}}|\Psi_{1,m'}\rangle$. It takes non-zero values only for interband  couplings between $\mathrm{LL}_0$ and $\mathrm{LL}_1$, but there is no optical couplings between orbitals inside the same LL. Without disorder, $m$ conincides with the OAM of the orbital, and $\Omega_{m',m}=\Omega_0(\ell)\delta_{m',m+\ell}$. With disorder, this still holds approximately for edge states, but not in the bulk, where arbitrary orbitals can be coupled.

The Hamiltonian for the LLs in graphene reads, $H_{\textrm{el}}=\sum_m \mu^{(0)}_m a_{m}^\dagger a_{m} +\mu^{(1)}_m b_{m}^\dagger b_{m}$. Here $\mu^{(0)}_m$ and $\mu^{(1)}_m$ are the energies of orbitals in $\mathrm{LL}_0$ and $\mathrm{LL}_1$ respectively. The total Hamiltonian is given as, $H=H_{\textrm{el}}+H_{\textrm{I}}$. We define the interband polarization as, $P_{m,m'}=\langle a^\dagger_{m'} b_{m}\rangle$ and intraband polarizations as $T^{(0)}_{m,m'}=\langle a^\dagger_{m'} a_{m}\rangle$ for $\mathrm{LL}_0$, $T^{(1)}_{m,m'}=\langle b^\dagger_{m'} b_{m}\rangle$ for $\mathrm{LL}_1$, as illustrated in Fig.~\ref{Fig1}(b). When $m=m'$, the intraband polarization equals the occupation $T^{(0/1)}_{m,m}=\rho^{(0/1)}_{m}$. 

From the Heisenberg equation of motion, we derive the coupled Bloch equations: for intraband polarizations of $\mathrm{LL}_0$,
\begin{eqnarray}
\dot{T}^{(0)}_{n,n'} &=&i\Delta^{(0)}_{n',n} T^{(0)}_{n,n'} + S^{(0)}_{n}(1-\rho^{(0)}_n)\delta_{n,n'} \nonumber \\ &+&i \displaystyle\sum_{m}(\Omega_{m,n'}P^*_{m,n} - \Omega^*_{m,n}P_{m,n'}),
  \label{Final_P00}
\end{eqnarray}
for intraband polarizations of $\mathrm{LL}_1$,
\begin{eqnarray}
\dot{T}^{(1)}_{n,n'} &=&i\Delta^{(1)}_{n',n} T^{(1)}_{n,n'} - S^{(1)}_{n}\rho^{(1)}_n \delta_{n,n'}, \nonumber \\ &-&i \displaystyle\sum_{m}(\Omega_{n,m}P^*_{n',m} - \Omega^*_{n',m}P_{n,m}),
 \label{Final_P11}
\end{eqnarray}
for interband polarizations,
\begin{eqnarray}
\dot{P}_{n,n'} &=& -i\Delta_{n,n'}P_{n,n'} - D\frac{1}{2}(S^{(0)}_{n'}+S^{(1)}_{n}) P_{n,n'} \nonumber \\
&-& i \displaystyle\sum_{m}(\Omega_{n,m}T^{(0)}_{m,n'} - \Omega_{m,n'}T^{(1)}_{n,m}),  \label{Final_P10}
\end{eqnarray}
within the rotating frame with respect to the laser frequency $\omega$, $\Delta^{(s)}_{n,n'}=\epsilon_{sn}-\epsilon_{sn'} $ and $\Delta_{n,n'}=\epsilon_{1n}-\Delta-\epsilon_{0n'} $. We choose $D=10$ to make the dephasing of the coherence much faster than the decay occupation lifetimes~\cite{haug2009quantum,Note1}.

We also include interband acoustic phonon relaxation~\cite{wendler2015ultrafast,wendler2014carrier,haug2009quantum}. In Eqns.~(\ref{Final_P00}-\ref{Final_P10}), $S^{(0)}_{n}$, $S^{(1)}_{n}$ are the scatter-in rate for $\mathrm{LL_0}$ and scatter-out rates for $\mathrm{LL_1}$ respectively: $S^{(0)}_{n}=\sum_{n'}\Gamma_{n,n'}\rho^{(1)}_{n'}$ and $S^{(1)}_{n}=\sum_{n'}\Gamma_{n',n}(1-\rho^{(0)}_{n'})$; $\Gamma_{n,n'}$ is the interband-polarization relaxation rate. The average relaxation time follows as $\tau=1/\langle\Gamma_{n,n'}\rangle$. See~\cite{Note1} for modeling details of the relaxation.

 Other relaxation mechanisms are not relevant in our case. For optical phonons in graphene, they are off-resonant with the energy gap. For Coulomb scatterings, we only excite carriers between the two lowest LLs, and other levels are either completely filled or empty, so Coulomb scattering is much slower than phonon scattering. Depending on the substrate, Coulomb scattering may become even slower due to screening.

\textit{Average radial position and the OAM-induced current.} Without disorder and confinement, and with initial polarizations set to zero, the Bloch equations, Eqns.~(\ref{Final_P00}-\ref{Final_P10}), reduce to a set of independent two-level systems.  Their exact solution yields a compact expression for the current in the weak excitation limit:
\begin{equation}
    I_r(\ell)= e \left( \frac{4|\Omega_0(\ell)|^4}{D^2\Gamma} \right)^{1/3}.
    \label{current_reference_approx}
\end{equation}
This is the maximum current one may get without disorder. Notably, it shows a novel scaling power of 1/3, which is a result of the Boltzmann scattering~\cite{haug2009quantum,wendler2014carrier,snoke1991carrier,snoke1992evolution,Note1}. For $\ell$ much smaller than the total number of orbitals considered, we can approximate $I_r(\ell)=I_r$, independent of $\ell$. We use $I_r$ as a reference scale in the following discussions.

Once we include disorder in the system, the Bloch equations are solved numerically, and we determine the current by evaluating the average radial position of electrons, $\langle r \rangle=Tr\{\hat{\rho}\hat{r}\}$. Semi-classically, we define the average current as,
$I= \sqrt{m^*}\frac{e}{l_{\textrm{c}}}  \frac{d\langle r\rangle}{dt}$. The factor $\sqrt{m^*}$ takes into account the circumference of the outer edge~\cite{Note1}.

\begin{figure}
\centering
\includegraphics[scale=1]{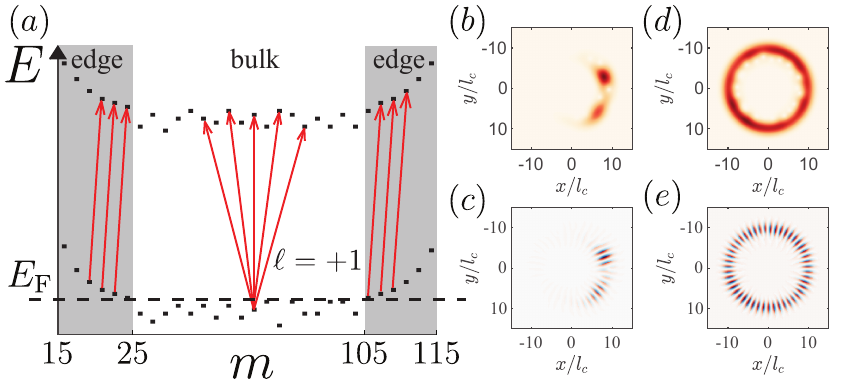}
\caption{(a) We illustrate the model used in the simulation. In total 100 orbitals are considered. The confinement is chosen such that 80 are bulk states while 10 are outer edge states and 10 are inner edge states. At $t=0$, the bulk states in $\mathrm{LL}_0$ are filled below the Fermi level $E_{\mathrm{F}}$, while others are empty. We use $\ell$=+1 to illustrate the optical selection rules. In the bulk, states can couple to many others, while on the edge only the states with OAM difference equal to $\ell$ are strongly coupled. (b) The wavefunction of a bulk state in real space is shown. The bulk state is localized, and the phase of the wavefunction is disordered, as shown in (c). (d) The edge state is delocalized and the twisted phase of the wavefunction (e) shows a well-defined OAM.}
\label{Fig2}
\end{figure}

\textit{Results.} In the simulation, we consider $m^*=100$ orbitals in each LL (Fig.~\ref{Fig2}(a)). In these orbitals, 80 of them are affected by the disorder in the bulk (Fig.~\ref{Fig2}(b)(c)) and become localized, whereas the 10 remain delocalized (Fig.~\ref{Fig2}(d)(e)) on the outer edge, and the other 10 are delocalized on the inner edge.
We set the Fermi energy such that initially the 80 orbitals in the bulk  are filled while the others are empty. We vary the average relaxation time $\tau$ from $50~\textrm{fs}$ to $10~\textrm{ps}$~\cite{plochocka2009slowing}. For the light beam, we choose the vortex to be located in the hole of the Corbino disk, such that on the disk the intensity profile can be assumed to be homogeneous, $A(r)=A_0$. In fact, spatial variations of the intensity profile away from the vortex do not affect the results. The vector potential is represented as $|A_0|=E_0/\omega$. Here, the electric field is chosen as $E_0=8.50\times10^5~{\rm V/m}$ which is accessible experimentally.

\begin{figure}
\centering
\includegraphics[scale=1]{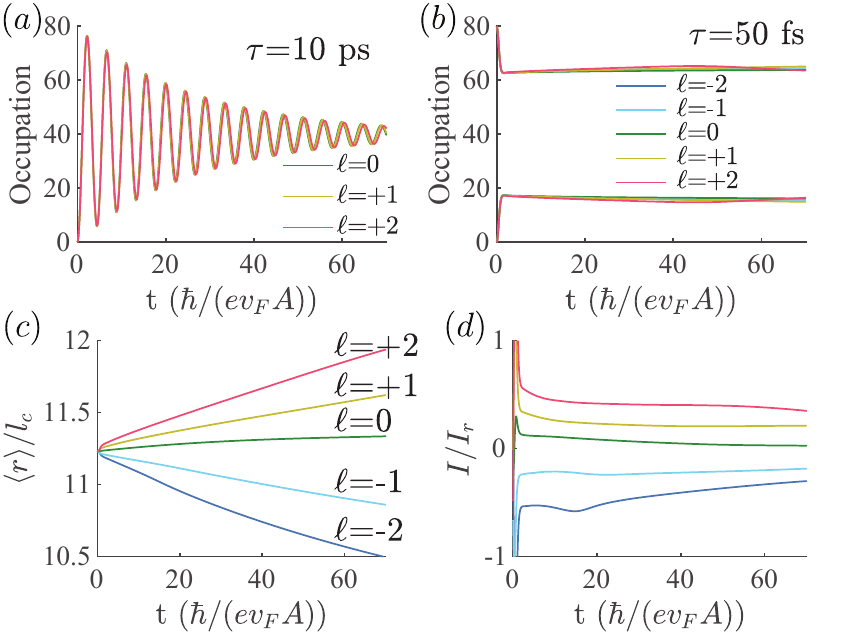}
\caption{(a) We plot the occupation of $\mathrm{LL}_1$ as a function of time for different $\ell$. We observe Rabi oscillations which are damped. (b) For the average relaxation time $\tau$=50~fs, one does not see Rabi oscillation because of the fast relaxation compared to the Rabi frequency. (c) The average radial position of electrons as a function of time for various OAM excitations, for $\tau=50~{\rm fs}$. (d) The semi-classical current corresponding to (c) for various OAM excitations.}
\label{Fig3}
\end{figure}

We turn on the continuous-wave laser and solve the Bloch equations of the system Eqns.~(\ref{Final_P00}-\ref{Final_P10}) to obtain the single particle density matrix $\hat{\rho}$ as a function of time. The average relaxation rate of electrons between LLs in graphene can vary depending on the magnetic field~\cite{plochocka2009slowing,mittendorff2015carrier}. For $\tau=10~\textrm{ps}$, we observe Rabi oscillations for different $\ell$'s, as shown in Fig.~\ref{Fig3}(a) where the occupation of $\mathrm{LL}_1$ is plotted as a function of time. These oscillations are quickly damped, which can be understood as a result of optically-induced diffusion. Specifically, light with OAM couples to orbitals which are distant from each other and therefore electrons diffuse with OAM excitation. Since orbitals have different couplings strength due to a random disorder, they have inhomogeneous Rabi frequencies. While electrons diffuse into disordered states, the number of the Rabi frequencies participating increases, and therefore the total oscillation is damped. For $\tau=50~\textrm{fs}$, we do not see significant Rabi oscillations, as shown in Fig.~\ref{Fig3}(b), because of the faster relaxation compared to the Rabi frequency.

We evaluate the observables $\langle r\rangle$ and $I$ as a function of time. In Fig.~\ref{Fig3}(c), we plot the average position of the 80 electrons as a function of time for excitations with different $\ell$. We find that for positive OAM, $\langle r\rangle$ increases with time, and larger OAM values lead to a faster increase. The resulting current is plotted in Fig.~\ref{Fig3}(d).

\begin{figure}
\centering
\includegraphics[scale=1]{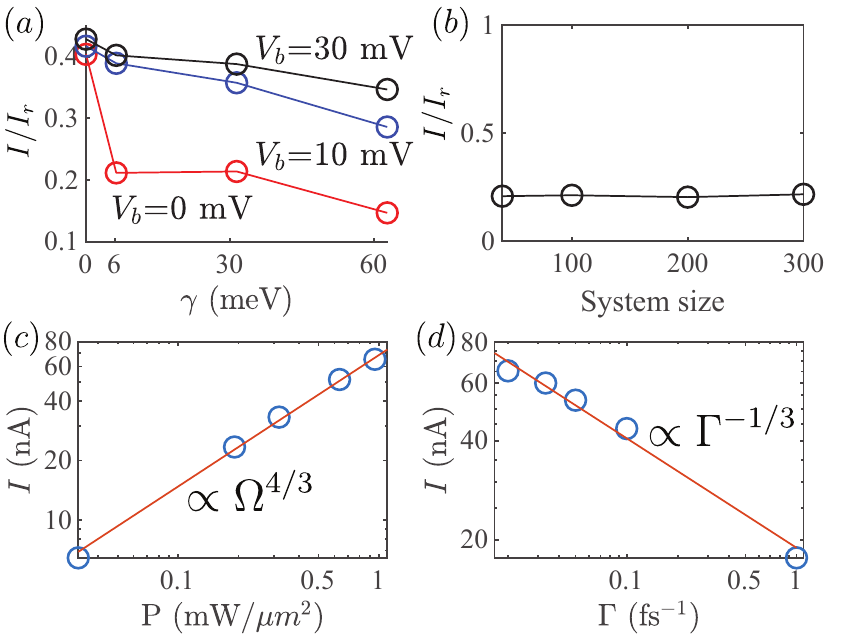}
\caption{(a) We plot the average current as a function of disorder strength $\gamma$, with various bias voltages $V_b$ =0, 10, 30 mV. Increasing $\gamma$ leads to a decrease of the average current. We apply a DC voltage bias $V_b$ across the sample to recover the current. We find a bias increases the average current. Here we use $\ell$=+1, $\tau$=50 fs. (b) We plot the average current for OAM=+1, $\tau$=50 fs as a function of total number of orbitals (system size) considered in the simulation. The average current stays constant. (c) We plot the simulated current $I$ as a function of pump intensity in log-log scales and compare with the reference current $I_r$ multiplied by a constant 0.21 (orange line). (d) We plot the simulated $I$ as a function of average relaxation rate $\Gamma=1/\tau$ in log-log scales and compare with the reference current $I_r$ multiplied by a constant 0.22 (orange line). The simulated results' scalings match very well with the analytical predictions.}
\label{Fig4}
\end{figure}

 We average the current from  Fig.~\ref{Fig3}(d) after it reaches equilibrium, and in Fig.~\ref{Fig4}(a) we plot this average as a function of disorder strength $\gamma$ for $\textrm{OAM}=+1$ and  $\tau=50~\textrm{fs}$. It shows that larger disorder diminishes the OAM-induced current. This is expected because stronger disorder may introduce couplings between more disorder eigenstates and invalidate the pairwise selection rules of the orbitals. However, we can recover the current from disorder through a voltage bias.

Voltage biases control the system properties through the DC-Stark effect \cite{empedocles1997quantum} or the Franz-Keldysh effect \cite{haug2009quantum}. Here, we use a DC voltage to bias the two contacts of the sample and induce a DC current through the sample. Simplistically, this is equivalent to adding a potential gradient onto the radial direction of the sample~\cite{price2014quantum,lukose2007novel,peres2007algebraic}. Experimentally, the DC bias does not interfere with the OAM-induced current because one may chop the laser and make the OAM-induced current alternating. Then the alternating signal may be picked out by using a frequency-locked lock-in amplifier, as a standard technique used in optoelectronics~\cite{gazzano2019observation,kasap2001optoelectronics}.

By having a small voltage bias across the Corbino sample, we can restore the rotational symmetry of the sample against disorders. In this way, the pairwise optical selection rules become valid again. As shown in Fig.~\ref{Fig4}(a), we plot the average current for voltage biases $V_b=10, 30~\textrm{mV}$ and without bias. Indeed, we see that voltage biases can recover the OAM-induced current.

In Fig~\ref{Fig4}(b), we plot the average current as a function of the total number of orbitals considered in the simulation (system size). It shows that the OAM-induced current is independent of the system size. In this simulation it is assumed that the coherence length always exceeds the system size. Indeed, it has been demonstrated that the coherence length in graphene in the quantum Hall regime can be as long as several $\mu m$ \cite{deprez2020tunable,ronen2020aharonov,wei2017mach,mills2019dirac}, comparable to the wavelength of the excitation.

Finally, we study the scaling of the OAM-induced current with pump intensity and relaxation. We have obtained an analytical expression for the current in Eqn.~(\ref{current_reference_approx}) for a disorder-free system in the weak pumping regime. In Fig.~\ref{Fig4}(c) and (d), we plot the simulated OAM-induced current $I$ for various pump intensities $P$ and average relaxation times $\tau$, respectively, in a disordered system. We compare with the scaling in Eqn.~(\ref{current_reference_approx}) and they match very well. Therefore, the scaling of the OAM-induced current is not affected by the disordered bulk.

\textit{Discussion.} We have proposed a measurement of the current resulting from the interactions between light with OAM and orbitals in LLs in graphene. We utilize the optical selection rules from the edge states, whose OAM is preserved due to the confinement potential, and adding a voltage bias extends the selection rules to even more states. The dynamics is an analogy to the Laughlin pump in the sense that flux insertion pumps charge through the system. In our scheme, however, this flux is added/removed by OAM of light through a non-adiabatic process, rather than adiabatically by a magnetic field. We find a scaling of the current with pump intensity to the power of 2/3, as a result of Pauli blockade. This result, analytically obtained for the system without disorder, also holds for disordered systems, as confirmed by numerical simulations. This coherent interplay with vortex light provides new strategies to probe and manipulate the topology of matter.

Not limited to graphene, similar effects could also be seen in other systems like conventional two-dimensional electron gas~\cite{girvin1999quantum}, where neglecting of Coulomb interactions is better justified due to strong screenings. On the other hand, Coulomb interactions and dynamical screenings~\cite{cooper1993coulomb} may play an important role in other materials, like transition metal dichalcogenide~\cite{cui2015multi,wu2016even,pisoni2018interactions}, and therefore, in the future, it will be interesting to study how the OAM-induced current is affected by Coulomb interactions. Furthermore, this idea of OAM-incuded current might be useful in probing the topology of the fast-developing field of twistronics, where correlated phases beyond LLs have been observed~\cite{das2021symmetry,wang2020correlated,shen2020correlated}.

\acknowledgments{The work in Maryland is supported by ARO W911NF2010232, ARL W911NF1920181, AFOSR FA95502010223, Simons foundation and the NSF funded PFC@JQI. T.G. acknowledges a fellowship granted by 
“la Caixa” Foundation (ID100010434, fellowship code LCF/BQ/PI19/11690013), and funding from  Fundació Privada Cellex, Fundació Mir-Puig, Generalitat de Catalunya (AGAUR Grant No. 2017 SGR 1341, CERCA program, QuantumCAT $\_$U16-011424, co-funded by ERDF Operational Program of Catalonia 2014-2020), Agencia Estatal de Investigación (“Severo Ochoa” Center of Excellence CEX2019-000910-S, Plan National FIDEUA PID2019-106901GB-I00/10.13039 / 501100011033, FPI), MINECO-EU QUANTERA MAQS (funded by State Research Agency (AEI) PCI2019-111828-2 / 10.13039/501100011033), EU Horizon 2020 FET-OPEN OPTOLogic (Grant No 899794), ERC AdG NOQIA, and the National Science Centre, Poland-Symfonia Grant No. 2016/20/W/ST4/00314.}

\bibliography{OAM_quantumHall_OpticalBloch}



\section*{Supplementary material}

# Supplemental Material for "Optical flux pump in the quantum Hall regime"

Bin Cao,[1] Tobias Grass,[2] Glenn Solomon,[1] and Mohammad Hafezi[1, 3]

[1]*Joint Quantum Institute, NIST/University of Maryland, College Park, Maryland 20742, USA*
[2]*ICFO-Institut de Ciencies Fotoniques, The Barcelona Institute of
Science and Technology, Castelldefels (Barcelona) 08860, Spain*
[3]*IREAP, University of Maryland, College Park, Maryland 20742, USA*

**CONTENTS**

## I. BLOCKED BEAM

We investigate the effect that part of the OAM beam is blocked by a contact bridge, as shown in Fig. S1(a). It is found that when the width of the bridge increases, the average OAM-incduced current slowly decreases as shown in Fig. S1(b).

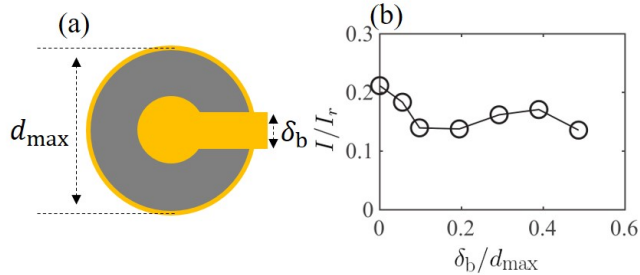

FIG. S1: (a) We study the scenario that there is a bridge contact on top across the Corbino sample, which blocks part of the OAM beam. The width of the bridge contact is $\delta_b$. (b) We plot the average OAM-induced current with OAM=+1, as a function of $\delta_b/d_{max}$ and the current slowly decreases.

## II. DISPLACED BEAM

We investigate the effect that the OAM beam's singularity is displaced from the center of the Corbino sample, as illustrated in Fig. S2(a). The average current sharply drops as a function of displacement $\delta_s$. A similar situation has been studied in [1] for atoms.

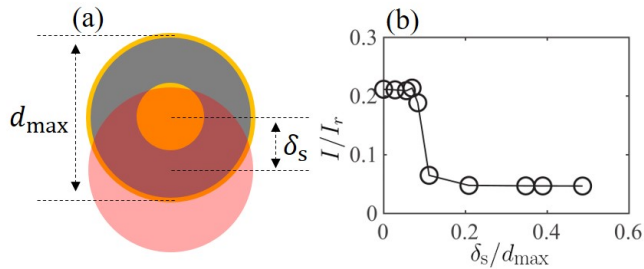

FIG. S2: (a) We study the effect that the OAM beam is shifted away from the center of the Corbino sample by a distance of $\delta_s$. (b) We plot the average OAM-induced current as a function of $\delta_s/d_{max}$ and find that the current drops sharply after a threshold.

## III. LIGHT-MATTER INTERACTION

The light matter interaction is obtained with the minimal coupling $\mathbf{p} \to \mathbf{p} - e\mathbf{A}$,

$$H_I(t) = ev_F \mathbf{A}(t) \cdot \sigma, \tag{S1}$$

where $\mathbf{A}(t)$ is the vector potential of light and it can be expressed as,

$$\mathbf{A}(t) = \mathbf{A}_0(r,\theta)e^{-i\omega t} + \mathbf{A}_0^*(r,\theta)e^{i\omega t}, \tag{S2}$$
$$\mathbf{A}_0(r,\theta) = A(r)e^{i\ell\theta}\mathbf{p}. \tag{S3}$$

Here, $A(r)$ is the mode of the light. In the rotating frame with frequency $\omega$, we remove the time-dependence and we have the light-matter interaction Hamiltonian in graphene as,

$$H_I = ev_F A(r)(e^{i\ell\theta}\sigma_- + e^{-i\ell\theta}\sigma_+) \tag{S4}$$

where we assume that the field is right-circular polarized, i.e. $\mathbf{p} = \mathbf{p}_+$.

We define the annihilation, creation operators $a_m, a_m^\dagger$ for electrons in orbital $m$ in $LL_0$ and $b_m, b_m^\dagger$ for $LL_1$. They satisfy, $\{a_m, a_{m'}^\dagger\} = \{b_m, b_{m'}^\dagger\} = \delta_{m,m'}$. We can rewrite the light-matter interaction as,

$$H_I = \sum_{m,m'} \Omega_{m',m} b_{m'}^\dagger a_m + \Omega_{m',m}^* a_m^\dagger b_{m'} \tag{S5}$$

The Rabi frequency for each pair of orbitals is obtained as, $\Omega_{m',m} = \langle \Psi_{0,m} | H_I | \Psi_{1,m'} \rangle$. We can also list some useful relations for $a$ and $a^\dagger$,

$$[a, a^\dagger a] = \{a, a^\dagger\}a = a \tag{S6}$$
$$[a^\dagger, a^\dagger a] = -a^\dagger\{a^\dagger, a\} = -a^\dagger, \tag{S7}$$

same relations apply for operators $b$ and $b^\dagger$.

## IV. OPTICAL BLOCH EQUATIONS

The Hamiltonian for the $LL_s$ in graphene reads, $H_{el} = \sum_m \mu_m^{(0)} a_m^\dagger a_m + \mu_m^{(1)} b_m^\dagger b_m$. Here $\mu_m^0$ and $\mu_m^1$ are the energies of orbitals in $LL_0$ and $LL_1$ respectively. The total Hamiltonian is given as,

$$H = H_{el} + H_I. \tag{S8}$$

We define the interband polarization as, $P_{m,m'} = \langle a_{m'}^\dagger b_m \rangle$ and intraband polarizations as $T_{m,m'}^{(0)} = \langle a_{m'}^\dagger a_m \rangle$ for LL$_0$, $T_{m,m'}^{(1)} = \langle b_{m'}^\dagger b_m \rangle$ for LL$_1$. We have the relation $T_{m,m'}^{(s),*} = T_{m',m}^{(s)}$. When $m = m'$, the intraband polarization is essentially the occupation $T_{m,m}^{(0/1)} = \rho_m^{(0/1)}$

The Heisenberg equation of motion is given as, $-i\frac{\partial}{\partial t}\hat{O} = [H, \hat{O}]$, where $\hat{O} = P_{m,m'}, T_{m,m'}^{(s)}$. From this, we derive the coupled optical Bloch equations.

### IV.1. Interband polarizations

As an example, we derive for the interband polarizations $P_{n',n}^* = \langle b_{n'}^\dagger a_n \rangle$. The commutator of $P_{n',n}^*$ with $H_{\text{el}}$ is,

$$\left[H_{\text{el}}, b_{n'}^\dagger a_n\right] = \sum_m \mu_m^0 \left[a_m^\dagger a_m, b_{n'}^\dagger a_n\right] + \sum_m \mu_m^1 \left[b_m^\dagger b_m, b_{n'}^\dagger a_n\right]$$
$$= -\mu_n^{(0)} b_{n'}^\dagger a_n + \mu_{n'}^{(1)} b_{n'}^\dagger a_n$$
$$= (\mu_{n'}^{(1)} - \mu_n^{(0)}) b_{n'}^\dagger a_n$$
$$= \Delta_{n',n} b_{n'}^\dagger a_n \tag{S9}$$

The commutator with $H_I$ is,

$$\left[H_I, b_{n'}^\dagger a_n\right] = \sum_{m,m'} \left[\Omega_{m',m} b_{m'}^\dagger a_m + \Omega_{m',m}^* a_m^\dagger b_{m'}, b_{n'}^\dagger a_n\right] \tag{S10}$$

The first term in Eqn. (S10) is,

$$\left[b_{m'}^\dagger a_m, b_{n'}^\dagger a_n\right] = b_{m'}^\dagger b_{n'}^\dagger a_m a_n - b_{n'}^\dagger b_{m'}^\dagger a_n a_m = 0. \tag{S11}$$

The second term in Eqn. (S10) is,

$$\left[a_m^\dagger b_{m'}, b_{n'}^\dagger a_n\right] = b_{m'} b_{n'}^\dagger a_m^\dagger a_n - b_{n'}^\dagger b_{m'} a_n a_m^\dagger$$
$$= (\delta_{m',n'} - b_{n'}^\dagger b_{m'}) a_m^\dagger a_n - b_{n'}^\dagger b_{m'} (\delta_{n,m} - a_m^\dagger a_n)$$
$$= a_m^\dagger a_n \delta_{m',n'} - b_{n'}^\dagger b_{m'} \delta_{n,m} \tag{S12}$$

Using Eqn. (S11), (S12) with Eqn. (S10) and combining with Eqn. (S9), we can arrive at the commutator with the total Hamiltonian $H$,

$$\left[H, b_{n'}^\dagger a_n\right] = \Delta_{n',n} b_{n'}^\dagger a_n + \sum_m (\Omega_{n',m}^* a_m^\dagger a_n - \Omega_{m,n}^* b_{n'}^\dagger b_m), \tag{S13}$$

i.e.,

$$\dot{P}_{n',n}^* = i\Delta_{n',n} P_{n',n}^* + i\sum_m (\Omega_{n',m}^* . T_{n,m}^{(0)} - \Omega_{m,n}^* T_{m,n'}^{(1)}). \tag{S14}$$

This gives the optical Bloch equation for the interband polarization, as given in the main paper but without the scattering terms.

### IV.2. Intraband polarization

As an example, we derive for LL$_1$ intraband polarizations. The intraband polarizations are $T_{n,n'}^{(1)} = \langle b_{n'}^\dagger b_n \rangle$. The commutator with $H_{\text{el}}$ is,

$$\left[H_{\text{el}}, b_{n'}^\dagger b_n\right] = \sum_m \mu_m^0 \left[a_m^\dagger a_m, b_{n'}^\dagger b_n\right] + \sum_m \mu_m^1 \left[b_m^\dagger b_m, b_{n'}^\dagger b_n\right]$$
$$= \sum_m \mu_m^1 b_{n'}^\dagger b_n (\delta_{m,n'} - \delta_{m,n})$$
$$= \Delta_{n',n}^{(1)} b_{n'}^\dagger b_n \tag{S15}$$

The commutator with $H_\mathrm{I}$ is,

$$\left[H_\mathrm{I},b^\dagger_{n'}b_n\right] = \sum_{m,m'}\left[\Omega_{m',m}b^\dagger_{m'}a_m + \Omega^*_{m',m}a^\dagger_m b_{m'}, b^\dagger_{n'}b_n\right]$$

$$= -\sum_{m,m'}\Omega_{m',m}b^\dagger_{n'}a_m\delta_{m',n} + \sum_{m,m'}\Omega^*_{m',m}a^\dagger_m b_n\delta_{n',m'}$$

$$= -\sum_m(\Omega_{n,m}b^\dagger_{n'}a_m - \Omega^*_{n',m}a^\dagger_m b_n) \quad (S16)$$

Note that in the derivation of Eqn. (S16), we have used the relations,

$$\left[b^\dagger_{m'}a_m, b^\dagger_{n'}b_n\right] = b^\dagger_{m'}a_m b^\dagger_{n'}b_n - b^\dagger_{n'}b_n b^\dagger_{m'}a_m$$

$$= (b^\dagger_{m'}b^\dagger_{n'}b_n - b^\dagger_{n'}b_n b^\dagger_{m'})a_m$$

$$= (b^\dagger_{m'}b^\dagger_{n'}b_n - b^\dagger_{n'}(\delta_{n,m'} - b^\dagger_{m'}b_n))a_m$$

$$= (\{b^\dagger_{m'},b^\dagger_{n'}\}b_n - b^\dagger_{n'}\delta_{n,m'})a_m$$

$$= -b^\dagger_{n'}a_m\delta_{n,m'}, \quad (S17)$$

and,

$$\left[a^\dagger_m b_{m'}, b^\dagger_{n'}b_n\right] = a^\dagger_m(b_{m'}b^\dagger_{n'}b_n - b^\dagger_{n'}b_n b_{m'})$$

$$= a^\dagger_m(\delta_{n',m'}b_n - b^\dagger_{n'}\{b_{m'},b_n\})$$

$$= a^\dagger_m b_n\delta_{n',m'} \quad (S18)$$

Combining Eqn. (S15) and Eqn. (S16), we arrive at,

$$\left[H,b^\dagger_{n'}b_n\right] = \Delta^{(1)}_{n',n}b^\dagger_{n'}b_n - \sum_m(\Omega_{n,m}b^\dagger_{n'}a_m - \Omega^*_{n',m}a^\dagger_m b_n), \quad (S19)$$

i.e.,

$$\dot{T}^{(1)}_{n,n'} = i\Delta^{(1)}_{n',n}T^{(1)}_{n,n'} - i\sum_m(\Omega_{n,m}P^*_{n',m} - \Omega^*_{n',m}P_{n,m}), \quad (S20)$$

which is the optical Bloch equation for intraband polarizations in $\mathrm{LL}_1$ as given in the main paper, without the scattering terms.

### IV.3. The optical Bloch equations

We add the Boltzmann scattering terms into the optical Bloch equations and arrive at the complete set of the optical Bloch equations, for intraband polarizations,

$$\dot{T}^{(0)}_{n,n'} = i\Delta^{(0)}_{n',n}T^{(0)}_{n,n'} + i\sum_m(\Omega_{m,n'}P^*_{m,n} - \Omega^*_{m,n}P_{m,n'}) + S^{(0)}_n(1-\rho^{(0)}_n)\delta_{n,n'}, \quad (S21)$$

$$\dot{T}^{(1)}_{n,n'} = i\Delta^{(1)}_{n',n}T^{(1)}_{n,n'} - i\sum_m(\Omega_{n,m}P^*_{n',m} - \Omega^*_{n',m}P_{n,m}) - S^{(1)}_n\rho^{(1)}_n\delta_{n,n'}, \quad (S22)$$

for interband polarizations

$$\dot{P}_{n,n'} = -i\Delta_{n,n'}P_{n,n'} - i\sum_m(\Omega_{n,m}T^{(0)}_{m,n'} - \Omega_{m,n'}T^{(1)}_{n,m}) - D\frac{1}{2}(S^{(0)}_{n'} + S^{(1)}_n)P_{n,n'}, \quad (S23)$$

To check the indices for the optical Bloch equations, we can start with the occupation conservation. For occupations, we take $n=n'$ and we sum over $n$, we can get the time derivative of the total occupation as,

$$\dot{\rho} = \sum_n(\dot{\rho}^{(0)}_n + \dot{\rho}^{(1)}_n)$$

$$= i\sum_{n,m}(\Omega_{m,n}P^*_{m,n} - \Omega^*_{m,n}P_{m,n}) - i\sum_{n,m}(\Omega_{n,m}P^*_{n,m} - \Omega^*_{n,m}P_{n,m}))$$

$$= 0. \quad (S24)$$

Thus, the total occupation is conserved indicating that the indices are correct.

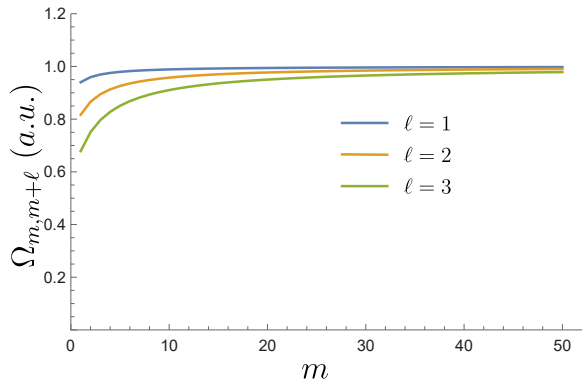

FIG. S3: We plot $\Omega_{m,m+\ell}$ as a function of $m$ for fixed $\ell$'s using Eqn. (S26). The coupling is approximately independent of $m$ for $m > 15$.

## V. ANALYTICAL ANALYSIS

### V.1. Translational symmetry

Without disorder or edge confinement, the orbitals in graphene LLs can be treated approximately as translational symmetric in the orbital index $m$. By translational symmetric, we mean the relaxation rates and the optical coupling (Rabi frequency) between orbitals do not depend on $m$. In order to justify this, we divide our following discussion into two parts: relaxation rates and optical couplings.

The relaxation rate is simple to be justified as not dependent on $m$. This is because the phonon relaxation rates are determined by the overlap of the wavefunctions [2]. In a system without any disorder or edge, the overlap of orbitals with different $m$ is always zero while with the same $m$, the overlap is a constant $1/\sqrt{2}$.

Does the optical coupling or the Rabi frequency $\Omega_{m,m+\ell}$ between $\Psi_{0,m+\ell}$ and $\Psi_{1,m}$, depend on $m$ or $\ell$? Physically, the possible dependence of $\Omega_{m,m+\ell}$ on $m$ and $\ell$ can be attributed to two aspects: 1, the twisted phase difference between the wavefunctions; 2, the spatial overlap of the wavefunctions in the real space controlled by $m$ and $\ell$. The phase difference in the first aspect is compensated by the additional OAM provided by the light and thus should not contribute. Here, we focus on the second aspect.

To take account of the spatial dependence, we calculate overlap of wavefunctions assuming the phase difference has been compensated. Specifically, we take $m + \ell$ orbital in $LL_0$ and $m$ orbital in $LL_1$,

$$\Psi_{0,m+\ell} = \begin{pmatrix} 0 \\ |0, m+\ell\rangle \end{pmatrix}, \Psi_{1,m} = \frac{1}{\sqrt{2}} \begin{pmatrix} |0, m\rangle \\ |1, m\rangle \end{pmatrix}. \tag{S25}$$

We calculate the couplings of the two wavefunctions, assuming the system is illuminated by a circularly polarized OAM=$\ell$ light with a homogeneous intensity,

$$\Omega_{m,m+\ell} = ev_{\rm F}\langle\Psi_{0,m+\ell}|\mathbf{A}(r)|\Psi_{1,m}\rangle = ev_{\rm F}\frac{A(r)}{\sqrt{2}}\frac{(-i)^\ell}{\sqrt{m!(m+\ell)!}}\Gamma(m + \frac{\ell}{2} + 1). \tag{S26}$$

With fixed $\ell$'s, we plot $\Omega_{m,m+\ell}$ as a function of $m$ in Fig. S3. It is found that if $m$ is large enough, the wavefunction overlap does not depend on the value of $m$ or $\ell$ anymore. In other words, $\Omega_{m,m+\ell}$ becomes independent of $m$ if $m$ is large enough. From Fig. S3, we can see the critical value for $m$ (for $\ell \leq 3$) is approximately 15. This means that $m$ needs to be larger than 15 in order to have the translational symmetry assumption valid.

On the other hand, we plot $\Omega_{m,m+\ell}$ as a function of $\ell$ with fixed $m$'s in Fig. S4. It is found that the overlap decays with $\ell$. This means that the optical coupling depends on the value of $\ell$. However, as shown in Fig. S4, the decay is slower for larger $m$. In the simulations, we consider $\ell < 2$ and $m$ between 15 and 115; thus we may ignore the dependence by using $\Omega_0(\ell) = \Omega_0$.

In conclusion, we find that it is valid to assume that the system is translational symmetric when $m$ is large. The Rabi frequency does not depend on $m$ when $m > 15$. On the other hand, since we consider small $\ell(\leq 2)$ for the excitation, we can drop the dependence of the Rabi frequency on $\ell$ as well, by simply having $\Omega_0(\ell) = \Omega_0$.

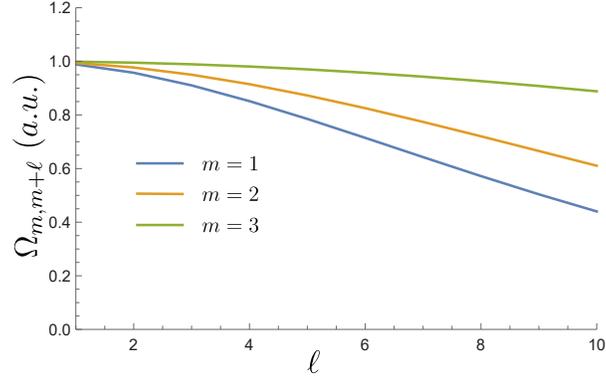

FIG. S4: We plot $\Omega_{m,m+\ell}$ as a function of $\ell$ for fixed $m$'s using Eqn. (S26). The coupling decreases with $\ell$. But for $\ell < 2$, we may approximate the coupling independent of $\ell$.

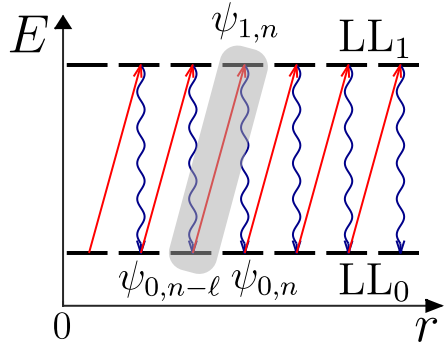

FIG. S5: An infinite system is divided into subsystems made of two levels, as shown in the shaded region.

### V.2. Reference current $I_r$

In this section, we show how we arrive at the analytical expression for the reference current $I_r$. We consider an ideal model without disorder, composed of an infinite number of orbitals in each LL, as shown in Fig. S5. In this context, we have translational symmetry in each LL, as explained in section V.1. We further assume that the OAM=$\ell$ beam is in resonance with the two LLs so the detuning terms are ignored.

Because of translational symmetry, at equilibrium we have $\rho_m^{(0)} = \rho_n^{(0)}$ and $\rho_m^{(1)} = \rho_n^{(1)}$ for any $n,m$. Thus, the intraband polarizations should be zero, $T_{m,n}^{(0)} = 0, T_{m,n}^{(1)} = 0$ for $n \neq m$. Also, only certain orbitals are optically coupled. In our case, we use OAM=$\ell$ so that $\Omega_{n,m} = \Omega_0 \delta_{n,m+\ell}$. On the other hand, the relaxation is pairwise as well. Specifically, the relaxation is only between orbitals with index $n$ in $LL_1$ and $n$ in $LL_0$.

Now we focus on the individual two-level element of the system. Specifically, we divide the infinite systems into elements which are made of two levels, as shown in the shaded region in Fig. S5. In the shaded region, we isolate two levels which as a whole is a building block of the total system. First we take a look at the interband polarizations, Eqn. (S23). Knowing that $T_{m,n}^{(0)} = 0, T_{m,n}^{(1)} = 0$ for $n \neq m$ and $\rho_m^{(0)} = \rho_n^{(0)}$, we can simplify Eqn. (S23) as,

$$\dot{P}_{n,n'} = -i(\Omega_{n,n'}\rho_{n'}^{(0)} - \Omega_{n,n'}\rho_n^{(1)}) - D\frac{1}{2}(S_{n'}^{(0)} + S_n^{(1)})P_{n,n'} \tag{S27}$$

Since the Rabi coupling is only between orbitals with OAM difference equal to $\ell$, i.e. $\Omega_{n,n'} = \Omega_0 \delta_{n,n'+\ell}$, we can further simplify Eqn. (S27) to,

$$\begin{aligned}
\dot{P}_{n+\ell,n} &= -i(\Omega_{n+\ell,n}\rho_n^{(0)} - \Omega_{n+\ell,n}\rho_{n+\ell}^{(1)}) - D\frac{1}{2}(S_n^{(0)} + S_{n+\ell}^{(1)})P_{n+\ell,n} \\
&= -i\Omega_{n+\ell,n}(\rho_n^{(0)} - \rho_{n+\ell}^{(1)}) - D\frac{1}{2}(\Gamma\rho_n^{(1)} + \Gamma(1-\rho_{n+\ell}^{(0)}))P_{n+\ell,n} \\
&= -i\Omega_{n+\ell,n}(\rho_n^{(0)} - \rho_{n+\ell}^{(1)}) - D\frac{1}{2}(\Gamma\rho_n^{(1)} + \Gamma(1-\rho_n^{(0)}))P_{n+\ell,n} \\
&= -i\Omega_{n+\ell,n}(2\rho_n^{(0)} - 1) - D\Gamma(1-\rho_n^{(0)})P_{n+\ell,n}.
\end{aligned} \tag{S28}$$

In the derivation, we have used the relation $S_n^{(0)} = \Gamma \rho_n^{(1)}$ and $S_n^{(1)} = \Gamma(1-\rho_n^{(0)})$. We also use the relation $\rho_{n+\ell}^{(1)} = \rho_n^{(1)}$, $\rho_{n+\ell}^{(0)} = \rho_n^{(0)}$, because of translational symmetry. In addition, $\rho_n^{(1)} = 1 - \rho_n^{(0)}$ as a result of occupation conservation and translational symmetry.

In the equilibrium, we have $\dot{P}_{n+\ell,n} = 0$ and with Eqn. (S28) we obtain,

$$P_{n+\ell,n} = \frac{-i\Omega_{n+\ell,n}(2\rho_n^{(0)} - 1)}{D\Gamma(1 - \rho_n^{(0)})} = \frac{i\Omega_{n+\ell,n}}{D\Gamma}(2 - \frac{1}{1 - \rho_n^{(0)}}) \tag{S29}$$

With this, we switch to the equation of occupations in a LL$_0$ orbital, derived from Eqn. (S21),

$$\begin{aligned}\dot{\rho}_n^{(0)} &= i(\Omega_{n+\ell,n}P^*_{n+\ell,n} - \Omega^*_{n+\ell,n}P_{n+\ell,n}) + S_n^{(0)}(1 - \rho_n^{(0)}) \\ &= -2\text{Im}(\Omega_{n+\ell,n}P^*_{n+\ell,n}) + \Gamma\rho_n^{(1)}(1 - \rho_n^{(0)})\end{aligned} \tag{S30}$$

We plug Eqn.(S29) into Eqn. (S30) and obtain,

$$\begin{aligned}\dot{\rho}_n^{(0)} &= -2\text{Im}(\Omega_{n+\ell,n}P^*_{n+\ell,n}) + \Gamma\rho_n^{(1)}(1 - \rho_n^{(0)}) \\ &= 2\frac{|\Omega_{n+\ell,n}|^2}{D\Gamma}(2 - \frac{1}{1 - \rho_n^{(0)}}) + \Gamma(1 - \rho_n^{(0)})^2 \\ &= 2\frac{|\Omega_0(\ell)|^2}{D\Gamma}(2 - \frac{1}{1 - \rho_n^{(0)}}) + \Gamma(1 - \rho_n^{(0)})^2,\end{aligned} \tag{S31}$$

where we change the notation for $\Omega_{n+\ell,n}$ to $\Omega_0(\ell)$ because of translational symmetry. We need to solve Eqn. (S31) to get the time evolution of $\rho_n^{(0)}$. We firstly take a qualitative look of Eqn. (S31). Using the typical parameters $\Omega_0(\ell) = 2\pi/(0.1 \text{ ps}), D = 10, \Gamma = 1/(50 \text{ fs})$ used in the paper, we can solve Eqn. (S31) numerically, and plot the occupation $\rho_n^{(0)}$ as a function of time in Fig. S6. The initial condition is $\rho_n^{(0)}(0) = 1$. For small $t$, the first term of Eqn. (S31) is predominant, giving a very large negative slope. As $\rho_n^{(0)}$ decreases, the absolute value of the first term becomes small and the second term becomes comparable to the first at the end of the dynamics. Eventually an equilibrium is reached.

Then we can solve for the equilibrium occupation analytically at $t \to \infty$. The result yields a complicated expression,

$$\rho_n^{(0)}(t \to \infty) = 1 - \frac{1}{3}\left(\frac{4\cdot 3^{2/3}A}{\sqrt[3]{\sqrt{3}\sqrt{A^2\Gamma^3(64A + 27\Gamma)} - 9A\Gamma^2}} - \frac{\sqrt[3]{3\sqrt{3}\sqrt{A^2\Gamma^3(64A + 27\Gamma)} - 27A\Gamma^2}}{\Gamma}\right), \tag{S32}$$

where $A = \frac{|\Omega_0(\ell)|^2}{D\Gamma}$.

When the equilibrium is reached, the two terms in Eqn. (S31) cancel each other and there is a persistent current along the radial direction. The first term in Eqn. (S31) is from optical couplings and the second term is the relaxation from the orbital right above. Since relaxation does not directly give rise to a radial current, our radial current is purely from the first term in Eqn. (S31). Specifically, the current at equilibrium is,

$$I_r(\ell) = 2e\frac{|\Omega_0(\ell)|^2}{D\Gamma}(\frac{1}{1 - \rho_n^{(0)}(t \to \infty)} - 2), \tag{S33}$$

where $e$ is the electron charge. $\rho_n^{(0)}(t \to \infty)$ is as given in Eqn. (S32). Note that, compared to the original term in Eqn. (S31), we add a minus sign in Eqn. (S33) because a positive current corresponds to a decrease of $\rho_n^{(0)}$.

We plot the equilibrium current $I_r$ as a function of $\Omega_0(\ell)$ and $\Gamma$ in Fig. S7. $D = 10$ is taken as a constant here. We can see the current increases with faster Rabi oscillation and faster relaxation.

In the weak pumping regime ($A \ll \Gamma$), we can simplify Eqn. (S32). The equilibrium occupation taken to the first order is,

$$\rho_n^{(0)}(t \to \infty) = 1 - \frac{2}{\sqrt[3]{4}}(\frac{A}{\Gamma})^{1/3} \tag{S34}$$

We can plug Eqn. (S34) in Eqn. (S33) and obtain a neat expression for the radial current in the limit $A \ll \Gamma$,

$$\begin{aligned}I_r(\ell) &= 2e\frac{|\Omega_0(\ell)|^2}{D\Gamma}(\frac{\sqrt[3]{4}}{2}(\frac{D\Gamma^2}{|\Omega_0(\ell)|^2})^{1/3} - 2) \\ &= e\left(\frac{4|\Omega_0(\ell)|^4}{D^2\Gamma}\right)^{1/3}\end{aligned} \tag{S35}$$

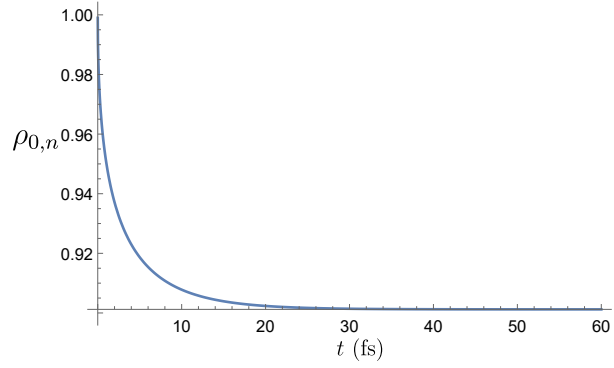

FIG. S6: We plot the occupation $\rho_{0,n}$ as a function of $t$, by numerically solving Eqn. (S31). An equilibrium is reached at the end.

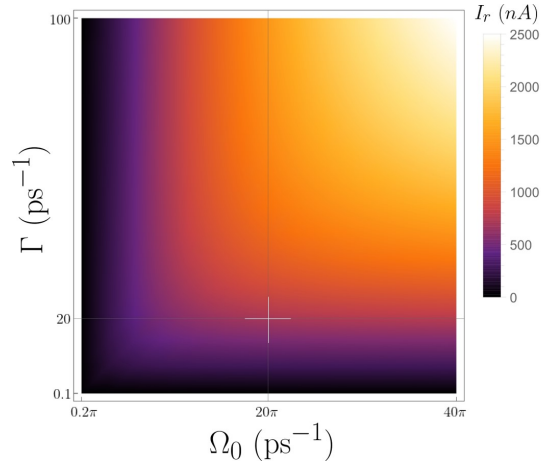

FIG. S7: We plot the current $I_r$ as a function of $\Omega_0$ and $\Gamma$, based on Eqn. (S33). $I_r$ increases with faster Rabi oscillation and faster relaxation. The parameters we use in most of our simulations: $\Omega_0(\ell) = 2\pi/(0.1 \text{ ps}), \Gamma = 1/(50 \text{ fs})$ are represented with the cross.

### V.3. Two level systems with conventional scattering

In order to get a deeper understanding of the electrons' behaviors, we simplify the Bloch equations shown above to describe a two level system (TLS). In this and the following sections, we solve the Bloch equations for a TLS with conventional and Boltzmann scatterings, respectively. Note that the OAM illuminated LLs cannot be considered as isolated TLSes and the full analysis has been given above. However from these two sections, we aim to understand where the novel scaling in Eqn. (S35) is from.

In this section, we simplify the Bloch equation to describe a TLS and use the relaxation rate from the conventional spontaneous scattering: $\Gamma\rho$ [3], instead of the Boltzmann scattering $\Gamma(1-\rho)\rho$. Specifically, the scattering rates are given as,

$$\begin{aligned}\dot{\rho}_{ee} &= -\Gamma\rho_{ee} \\ \dot{\rho}_{gg} &= \Gamma\rho_{ee} \\ \dot{\rho}_{eg} &= -\frac{\Gamma}{2}\rho_{eg}.\end{aligned} \quad (S36)$$

In order to describe a TLS, we simplify Eqn. (S28) and (S30) to the following,

$$\dot{\rho}_{eg} = -i\Omega_{eg}(2\rho_{gg} - 1) - \frac{\Gamma}{2}\rho_{eg} \quad (S37)$$

$$\begin{aligned}\dot{\rho}_{gg} &= i(\Omega_{eg}\rho_{eg}^* - \Omega_{eg}^*\rho_{eg}) + \Gamma\rho_{ee} \\ &= -2\text{Im}(\Omega_{eg}\rho_{eg}^*) + \Gamma(1 - \rho_{gg}),\end{aligned} \quad (S38)$$

where the notation $\Omega_{n+\ell,n}$ in Eqn. (S28) and (S30) has been changed to $\Omega_{eg}$. We obtain the equilibrium by setting $\dot{\rho}_{eg} = 0$, $\dot{\rho}_{gg} = 0$,

$$\rho_{eg} = -\frac{2i\Omega_{eg}}{\Gamma}(2\rho_{gg}(t \to \infty) - 1) \tag{S39}$$

$$\rho_{gg}(t \to \infty) = 1 - \frac{2\text{Im}(\Omega_{eg}\rho_{eg}^*)}{\Gamma}. \tag{S40}$$

In the weak excitation limit $\Omega_{eg} \ll \Gamma$, we use the first term in Eqn. (S38) combining with Eqn. (S39), (S40) to arrive at the "equilibrium current" for a TLS,

$$I_r^{\text{TLS}} = 4e\frac{|\Omega_{eg}|^2}{\Gamma}, \tag{S41}$$

which is intuitively consistent with a TLS in terms of scaling.

In summary, we replace the Boltzmann scatterings in the Bloch equations for OAM illuminated LLs with conventional scatterings, and arrive at the same scaling for a conventional TLS. This hints that the difference in the scalings of Eqn. (S35) and (S41) originates from distinct relaxations. To corroborate this, in the next section we use a conventional TLS but with Boltzmann scattering terms to replicate the novel scaling.

### V.4. Two level systems with Boltzmann scattering

The OAM illuminated LLs should not be considered as isolated TLSes and the full analysis is given in sections V.1 and V.2. However in order to capture some of the electron behaviors in OAM illuminated LLs, we modify the Bloch equations for a TLS with Boltzmann scattering terms,

$$\dot{\rho}_{gg} = i(\Omega_{eg}\rho_{eg}^* - \Omega_{eg}^*\rho_{eg}) + \Gamma\rho_{ee}(1 - \rho_{gg}) \tag{S42}$$

$$\dot{\rho}_{ee} = -i(\Omega_{eg}\rho_{eg}^* - \Omega_{eg}^*\rho_{eg}) - \Gamma\rho_{ee}(1 - \rho_{gg}) \tag{S43}$$

$$\dot{\rho}_{eg} = -i\Omega_{eg}(\rho_{gg} - \rho_{ee}) - \frac{1}{2}\Gamma(\rho_{ee} + 1 - \rho_{gg})\rho_{eg} \tag{S44}$$

In the equilibrium, Eqn. (S42)-(S44) are all equal to 0. Since $\rho_{ee} = 1 - \rho_{gg}$, Eqn. (S44) gives,

$$\rho_{eg} = \frac{i\Omega_{eg}}{\Gamma}(2 - \frac{1}{1 - \rho_{gg}}). \tag{S45}$$

Plug Eqn. (S45) into (S42), we obtain,

$$\dot{\rho}_{gg} = \frac{2|\Omega_{eg}|^2}{\Gamma}(2 - \frac{1}{1 - \rho_{gg}}) + \Gamma(1 - \rho_{gg})^2. \tag{S46}$$

We can see Eqn. (S46) is the same as Eqn. (S31). Therefore, the 1/3 scaling appeared in Eqn. (S35) also applies to a TLS with Boltzmann scattering terms. Here, we demonstrate that the novel scaling is not due to the degenerate orbitals in LLs, but rather due to the Boltzmann scattering terms.

### V.5. Inhomogeneous electron density with radius

In the analysis in sections V.2-V.4, we use a translational symmetric model. Even though the system is translational symmetric in terms of $m$, it is not in terms of radius $r$ since the orbital radius $r_m \propto \sqrt{m}$, as shown in Fig. (S8). Here, we show that the current given in Eqn. (S35) is not affected and it is conserved over the sample.

The definition of current is the electron density times the average radial position changing rate. The electron density is not a constant, but rather it is dependent on $m$. Therefore, we have,

$$I_r = e\frac{d\rho}{dr} \cdot \frac{dr}{dt} = e\frac{d\rho}{dt}, \tag{S47}$$

where $d\rho/dr$ is the carrier density with $r$ and $dr/dt$ is the changing rate of the electrons' radial position; $dr = 2(\sqrt{m+1} - \sqrt{m})l_c$ is the spacing between the adjacent orbitals but it's canceled out in the final expression of the equilibrium current. As a result, the current as in Eqn. (S35) is conserved over the sample, independent of the index of the orbital $m$ or the radial position $r$. In the simulations where we consider disorder, we take into account this inhomogeneous electron density with radius by adding a factor $\sqrt{m^*}$, as we will discuss in section VI.3.

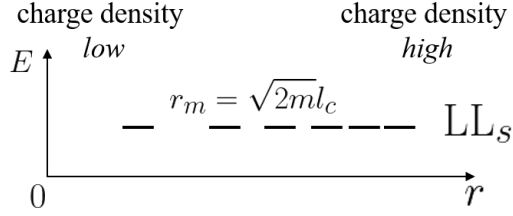

FIG. S8: In real space, the orbitals locate closer to each other as $m$ increases. This makes the charge density per radius increases with $m$.

## VI. SIMULATION

### VI.1. Effective edge

The system size in the simulation is limited by $m^*$. It corresponds to the maximum index of the possibly occupied orbitals, which is also close to the number of electrons considered in the simulation. Our simulation is valid until a point where the electrons arrive on the maximum orbital $m^*$. Therefore, we stop the simulation before a significant amount of electrons accumulate near the edge.

We simulate the edge using an effective potential. In particular, the effective edge confinement potential is a step function, since it has been observed that the rising potential on the edge is very sharp [4]. The details of the edge states may depend on the type of the edge [5]. Specifically, the edge could be zigzag or armchair, or gate-defined. However, the details of the edge should not affect the results because what is necessary in order to have the OAM-induced current is to have the edge states lifted in energy, from the disordered bulk.

The effective confinement potential for $LL_0$ depends on the valley index [6]. In particular, the potential for the $K$ valley is $V_c$ while for the $K'$ valley, it is $-V_c$. This gives rise to two branches of edge states for $LL_0$. However, electrons' behaviour of moving outward/inward under an OAM illumination does not depend on valley index since edge states from both valleys have defined OAMs.

In the simulation, we do not consider screening. The effects of screening on the edges might give rise to compressible and imcompressible stripes [4, 7, 8]. We want to note that this does not affect the twisted phase of the delocalized edge states and our conclusions are still valid.

### VI.2. Relaxation

We consider acoustic phonon relaxation in the simulation and ignore other relaxation mechanisms. The details of the phonon relaxation are given as in [2, 9]. The carrier phonon relaxation is described by the Hamiltonian,

$$H_{phon} = \sum_{if\mathbf{p}\mu} (g_{if}^{\mathbf{p}\mu} a_f^\dagger a_i b_{\mathbf{p}\mu} + g_{if}^{\mathbf{p}\mu *} a_i^\dagger a_f b_{\mathbf{p}\mu}^\dagger), \tag{S48}$$

where $\mu$ is the mode of the phonon. The carrier phonon matrix element with momentum $\mathbf{p}$ is given as,

$$g_{if}^{\mathbf{p}\mu} = \int d\mathbf{r}\, \Psi_f^*(\mathbf{r}) V_{phon}^\mu(\mathbf{p}) \Psi_i(\mathbf{r}). \tag{S49}$$

The coupling potential is,

$$V_{phon}^\mu(\mathbf{p}) = \begin{pmatrix} V_1 & V_2 \\ V_2^* & V_1 \end{pmatrix} = ipD_{\mathbf{p}}^\mu \begin{pmatrix} g_1 \cos\psi^- P & g_2 e^{-i\xi\psi^+} P \\ -g_2 e^{i\xi\psi^+} P & g_1 \cos\psi^- P \end{pmatrix}, \tag{S50}$$

where $V_1$ is the scalar deformation potential and $V_2$ is the modulated hopping due to the phonons. $g_1 = 16$ eV and $g_2 = -1.5$ eV. $P = e^{i\mathbf{p}\mathbf{R}}$ is the plane wave factor. $D_{\mathbf{p}}^\mu = \sqrt{\hbar/MA\omega_{\mathbf{p},\mu}}$, where the mass density $M = 7.6 \times 10^{-8}$ gcm$^{-2}$ and A is the area of graphene. In the long wavelength limit, we can ignore the plane wave factor $P$ in the coupling potential.

For clean orbitals without disorder, as in Eqn. (S25), we can get the carrier phonon matrix element, from Eqn. (S49) as,

$$|g_{if}^{\mathbf{p}\mu}| = |pD_{\mathbf{p}}^\mu \int d\mathbf{r}\, \frac{1}{\sqrt{2}} V_2^*| = |pD_{\mathbf{p}}^\mu \frac{A}{\sqrt{2}} V_2|. \tag{S51}$$

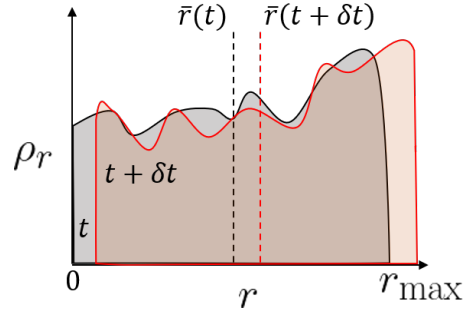

FIG. S9: Illustration of the calculation of current. At time $t$, the radial carrier distribution is shown as the gray-shaded region. It has a average radius of $\bar{r}(t)$. After an interval $\delta t$, the distribution is changed to the red-shaded region, which has an average radius $\bar{r}(t+\delta t)$.

For the disordered orbitals, we plug in the disordered wavefuntions shown as in Eqn. (2) in the main text, the carrier phonon matrix elements become,

$$|g_{n,n'}^{\boldsymbol{p}\mu,dis}| = |pD_{\boldsymbol{p}}^{\mu}\frac{A}{\sqrt{2}}V_2\sum_{\bar{m}}c_{1,\bar{m}}^{n',*}c_{0,\bar{m}}^{n}| \tag{S52}$$

For an orbital in $LL_1$, the total scattering rate is a sum over all the possible phonon momentum $\boldsymbol{p}$,

$$\begin{aligned}\Gamma_{n,n'} &= \frac{2\pi}{\hbar}\sum_{\boldsymbol{p}}|g_{n,n'}^{\boldsymbol{p}\mu,dis}|^2\delta(\Delta E) = \frac{8\pi^3}{A\hbar}\int d\boldsymbol{p}|g_{n,n'}^{\boldsymbol{p}\mu,dis}|^2\delta(\Delta E) \\ &= \frac{16\pi^4}{A\hbar}\int dp\,p|g_{n,n'}^{\boldsymbol{p}\mu,dis}|^2\delta(\Delta E).\end{aligned} \tag{S53}$$

Note that, $\frac{A}{(2\pi)^2}\sum_{\boldsymbol{p}} = \int d\boldsymbol{p} = \int d\phi \int dp\,p$. We use $\delta(\Delta E)$ to represent the broadenings of the individual orbitals within LLs.

In the integrand of Eqn. (S53), $|g_{n,n'}^{\boldsymbol{p}\mu,dis}|$ is dependent on $\omega_{\boldsymbol{p},\mu}$, which is the dispersion of phonons, through $D_{\boldsymbol{p}}^{\mu}$. Suppose we are in the low energy regime where the dispersion of acoustic phonons can be approximated by,

$$\omega_{\boldsymbol{p},\Gamma A} = \nu_{\boldsymbol{p},\Gamma A}|\boldsymbol{p}|, \tag{S54}$$

where $\nu_{\boldsymbol{p},\Gamma A}$ is the phonon velocity. Within this approximation, we can have Eqn. (S53) as,

$$\begin{aligned}\Gamma_{n,n'} &= \frac{8\pi^4}{M\nu_{\boldsymbol{p},\Gamma A}}g_2^2|\sum_m c_{0,\bar{m}}^n c_{1,\bar{m}}^{n',*}|^2\int dp\,p^2\delta(\Delta E) \\ &= \frac{8\pi^4}{M\hbar\nu_{\boldsymbol{p},\Gamma A}^2}g_2^2|\sum_m c_{0,\bar{m}}^n c_{1,\bar{m}}^{n',*}|^2 p_0^2 \\ &= \Gamma_0|\langle\Psi_{1,n'}|\Psi_{0,n}\rangle|^2\end{aligned} \tag{S55}$$

Here, $p_0 = (\epsilon_n - \epsilon_n')/\nu_{\boldsymbol{p},\Gamma A}$ is the momentum corresponding to the energy difference between the two orbitals $n$ and $n'$. $\Gamma_0 = \frac{8\pi^4}{M\hbar\nu_{\boldsymbol{p},\Gamma A}^2}g_2^2 p_0^2$ is a constant dependent on the intrinsic properties of the sample. Accordingly, we define the average relaxation time as $\tau = 1/\langle\Gamma_{n,n'}\rangle$. In the simulation, we vary $\tau$ between 50 fs and 10 ps.

If considering the Pauli exclusion, we add the Boltzmann scattering rate between states $|1,n'\rangle$ and $|0,n\rangle$ as,

$$S_n^{(0)} = \sum_{n'}\Gamma_{n,n'}\rho_{n'}^{(1)} \tag{S56}$$

$$S_{n'}^{(1)} = \sum_n \Gamma_{n,n'}(1-\rho_n^{(0)}) \tag{S57}$$

### VI.3. Current

The current of our interest is induced by the OAM of light. It has been demonstrated that an optical field can induce ultra-fast currents [10, 11]. But in our case, the current is different. Specifically, the current is composed of two

ingredients: excitation and relaxation. When the selection rules are valid, the excitation couples orbitals with different OAMs and equivalently, the electrons are transported spatially along the radial direction. The relaxation brings down the electrons from high energy, making the movement of electrons continuous. In the meantime, other current sources (photovoltaics, drifting, diffusion, etc.) are greatly suppressed due to the orthogonality of the wavefunctions.

We illustrate the calculation of current in Fig. S9. Suppose the distribution of carriers in the sample at time $t$, as illustrated in the gray-shaded region in Fig. S9, has an average radius $\bar{r}(t)$. After a small interval $\delta t$, the distribution becomes as shown in the red-shaded region with an average radius of $\bar{r}(t+\delta t)$. In the Corbino structure, the current is expressed as,

$$\begin{aligned} I &= \rho_{2D} v C \\ &= \frac{e}{\pi l_c^2} \frac{\bar{r}(t+\delta t) - \bar{r}(t)}{\delta t} 2\pi \sqrt{2\tilde{m}} l_c \\ &= 2\sqrt{2\tilde{m}} \frac{e}{l_c} \frac{d\langle r \rangle}{dt}, \end{aligned} \quad (S58)$$

where $C = 2\pi \bar{r}(t)$ is the circumference of the orbital $\tilde{m}$ which corresponds to the the average radius $\bar{r}(t) = \sqrt{2\tilde{m}} l_c$; $v = (\bar{r}(t+\delta t) - \bar{r}(t))/\delta t$ is the group velocity of the electrons; $\rho_{2D} = e/(\pi l_c^2)$ is the carrier density in a 2D system. Since $m^*$ is the characteristic index of the system size, we can rewrite the current Eqn. (S58) up to a constant as, $I = \sqrt{m^*} \frac{e}{l_c} \frac{d\langle r \rangle}{dt}$.

---



\end{document}